\documentclass[english,aps,prl,twocolumn,floats]{revtex4}
\usepackage[T1]{fontenc}
\usepackage[latin9]{inputenc}
\usepackage{amsmath}
\usepackage{amssymb}
\usepackage{graphicx}
\usepackage{esint}

\makeatletter

\@ifundefined{textcolor}{}
{%
 \definecolor{BLACK}{gray}{0}
 \definecolor{WHITE}{gray}{1}
 \definecolor{RED}{rgb}{1,0,0}
 \definecolor{GREEN}{rgb}{0,1,0}
 \definecolor{BLUE}{rgb}{0,0,1}
 \definecolor{CYAN}{cmyk}{1,0,0,0}
 \definecolor{MAGENTA}{cmyk}{0,1,0,0}
 \definecolor{YELLOW}{cmyk}{0,0,1,0}
}

\usepackage{babel}
\usepackage{bm}

\makeatother

\usepackage{babel}

\begin{document}

\title{Skyrmion Mass from Spin-Phonon Interaction}

\author{D. Capic, E. M. Chudnovsky, and D. A. Garanin}

\affiliation{Physics Department, Herbert H. Lehman College and Graduate School, The City University of New York \\
 250 Bedford Park Boulevard West, Bronx, New York 10468-1589, USA}

\date{\today}
\begin{abstract}
Inertial mass of a skyrmion arising from spin-phonon interaction is computed exactly within a toy model of the magnetoelastic coupling in a ferromagnetic film. The mass scales as the square of the strength of the magnetoelastic coupling, as the square of the film thickness, and as the first power of the lateral size of the skyrmion. For nanometer skyrmions it is in the ballpark of a few electron masses but may be significantly greater in materials with large magnetostriction. These findings are expected to stand for any complex structure of spin-phonon interaction in real materials. They must be taken into account when addressing the speed of information processing based upon skyrmions. 
\end{abstract}

\maketitle
Magnetic skyrmions are swirls of magnetization in thin films. They proliferated into condensed matter physics \cite{BelPolJETP75,WriMerRMR89,SonKarKivPRB93,StonePRB93,Bogdanov94,YeKimPRL99, AlkStoNat01} from field models of atomic nuclei and topologically stable elementary particles \cite{SkyrmePRC58,Polyakov-book,Manton-book,D1,D2}. In ferro- and antiferromagnets they are topological defects of the uniform magnetization (N\'{e}el vector) that cannot be easily destroyed. Unlike micron-size magnetic bubbles studied in the past \cite{MS-bubbles,ODell}, skyrmions can be small compared to the domain wall thickness, making them promising candidates for topologically-protected nanoscale information processing  \cite{Nagaosa2013,Zhang2015,Klaui2016,Leonov-NJP2016,Hoffmann-PhysRep2017,Fert-Nature2017}. 

Skyrmions can be  moved by current-induced spin-orbit torques \cite{Yu-NanoLet2016,Fert-Nature2017,Legrand-Nanolet2017, review2020}. The speed of the information processing with skyrmions depends on their inertia. The effort to compute and measure skyrmion inertial mass has been limited so far. The mass of a skyrmion bubble of dipolar origin, similar to the D$\ddot{\text{o}}$ring mass \cite{Doring} of the domain wall in the thin-wall approximation, has been discussed by Makhfudz et al. \cite{Makhfudz-PRL2012}. Large inertia has been reported in experiments on skyrmion breathing modes in the gigahertz frequency range \cite{Buttner-Nature2015}. Similar effects have been observed in the breathing and hypocycloid motion of skyrmions by Shiino et al. \cite{Shiino2017}. Inertial mass of electromagnetic origin due to excitation of magnons by a moving skyrmion has been studied by Lin \cite{Lin-PRB2017}. Psaroudaki et al. \cite{Psaroudaki-PRX2017} have demonstrated that translational symmetry makes classical skyrmions massless. They computed the mass arising from defects, non-uniformity of the magnetic field, and confining potentials, and elucidated contribution of thermal and quantum fluctuations to the mass.  Kravchuk et al. \cite{Kravchuk-PRB2018} used collective coordinates to demonstrate that skyrmion dynamics in a continuous spin-field model is massless even if one accounts for magnon excitations. Massive skyrmions have been reported by Li et al. \cite{Li-PRB2018} in simulations of collective magnetic dynamics on a two-dimensional honeycomb lattice. Measurement of the mass of the oscillating skyrmion in a confined geometry of a semicircular nano-ring has been recently proposed by Liu and Liang \cite{Liu-MMM2020}. 

The range for the skyrmion mass obtained in experiments is rather broad. It is not always clear whether it is associated with the confined geometry or more fundumental effects studied by theorists. The latter hint towards zero skyrmion mass in the presence of full translational invariance. Crystal lattice violates such invariance. In this Letter we show that skyrmions acquire a finite mass due to the spin-phonon interaction even within translationally invariant continuous spin-field and elastic theories. The physics behind the contribution of the atomic lattice to the skyrmion mass is transparent. The time-dependent spin field corresponding to the moving skyrmion induces, through the magnetoelastic coupling, the motion of the atoms whose inertia contributes to the mass of the skyrmion. Materials that host skyrmions are rather complex. In addition to the dominant exchange interaction, they possess various other kinds of magnetic interactions that are important for stabilization of skyrmions, such as Dzyaloshinskii-Moriya, Zeeman, and crystal field interactions \cite{Bogdanov94,Bogdanov-Nature2006,Heinze-Nature2011,Boulle-NatNano2016,Leonov-NJP2016}. Pertinent to our purpose, we shall consider in this Letter a toy model of Belavin-Polyakov skyrmion \cite{BelPolJETP75} interacting with isotropic elastic environment. We shall study two simple forms of the magnetoelastic coupling, the extreme anisotropic and fully isotropic. 

\begin{figure}
\includegraphics[width=170mm]{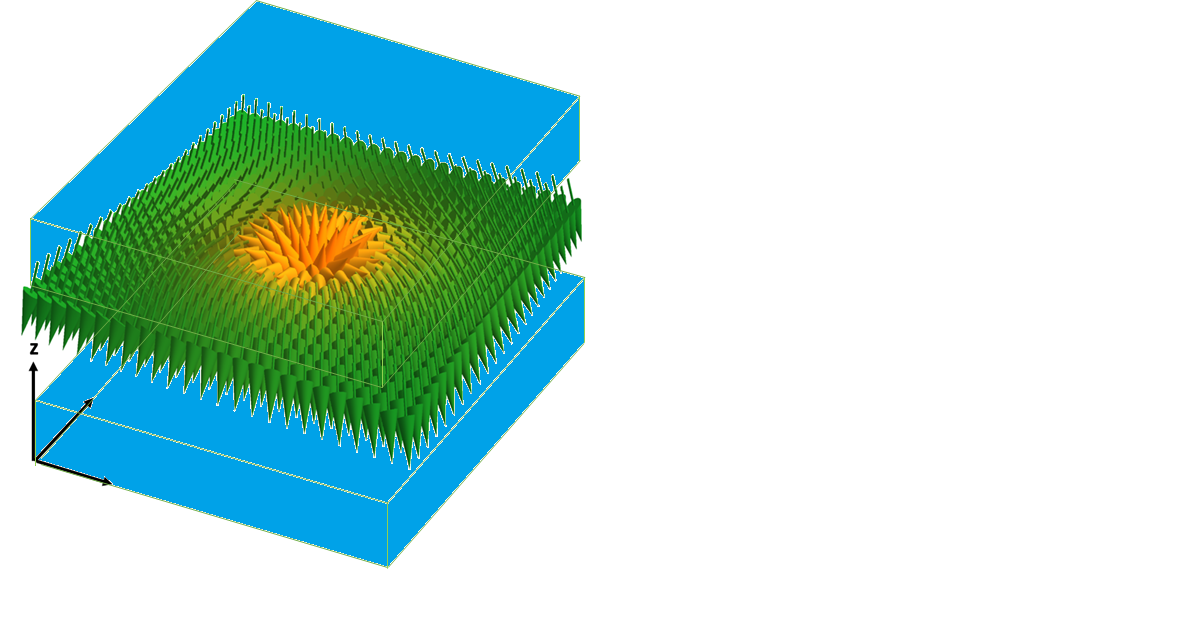}
\vspace{-8mm}
\caption{Color online: Skyrmion in a magnetic layer confined between two non-magnetic solids. Arrows show directions of the magnetization.}
\label{Fig}
\end{figure}
The skyrmion is modelled by the dimensionless three-component spin field ${\bf S}$ of unit length given by \cite{BelPolJETP75}
\begin{eqnarray}
S_x &  = & \frac{2\lambda(x\cos\gamma - y\sin\gamma)}{\lambda^2 + x^2 + y^2},\,\,  S_y = \frac{2\lambda(x\sin\gamma + y\cos\gamma)}{\lambda^2 + x^2 + y^2}, \nonumber \\
S_z & = & \frac{\lambda^2 - x^2 - y^2}{\lambda^2 + x^2 + y^2}, \quad {\bf S}^2 = S_x^2 + S_y^2 + S_z^2 = 1, \label{BP}
\end{eqnarray}
confined to the $xy$ layer of thickness $d$, with ${\bf S}$ looking down at infinity. Here $\lambda$ can be viewed as the lateral size of the skyrmion and $\gamma$ describes the rotation of the spin field, with $\gamma = 0,\pi$ and $\gamma = \pm\pi/2$ corresponding to the N\'{e}el-type and Bloch-type skyrmions respectively, see, e.g., 
Ref.\ \onlinecite{DCG-PRB2018}. Magnetoelastic interaction would generally be of the form $A_{ikjl}u_{ik}S_jS_l$, where 
\begin{equation}
u_{ik} = \frac{1}{2}\left(\frac{\partial u_i}{\partial r_k} + \frac{\partial u_k}{\partial r_i}\right)
\end{equation}
is the strain tensor, ${\bf u}$ is the phonon displacement field, and tensor $A_{ikjl}$ represents components of the magnetoelastic energy density.

We shall start with the extreme anisotropic form of the magnetoelastic coupling, $A u_{zz} S_z^2$, that together with the elastic contribution yields for the energy
\begin{small}  
\begin{equation}
E = \int d^3r \left[ \frac{1}{2}\rho \left(\frac{\partial {\bf u}}{\partial t}\right)^2 + \mu\left(u_{ik}^2 + \frac{\sigma}{1-2\sigma}u_{ll}^2\right) + A u_{zz} S_z^2\right]. \label{E-anisotropic}
\end{equation}
\end{small} 
To simplify the problem we assume that the magnetic layer is confined between two non-magnetic semi-infinite solids (see Fig.\ \ref{Fig}) having the same mass density $\rho$, the same shear modulus $\mu > 0$ and the same Poisson coefficient $\sigma = E/(2\mu) - 1$ (satisfying $-1 \leq \sigma \leq 1/2$), with $E$ being the Young modulus \cite{Elasticity}. If the speed of the moving skyrmion is small compared to the speed of sound, the elastic deformation adiabatically follows the skyrmion via extremal equation for the energy:
\begin{equation}
{\bm \nabla}^2{\bf u} + \frac{1}{1-2\sigma}{\bm \nabla}({\bm \nabla} \cdot {\bf u}) = -\frac{A}{\mu}\frac{\partial S_z^2}{\partial z}{\bf e}_z, \label{eq-u-an}
\end{equation}
where ${\bf e}_z$ is the unit vector along the $z$-axis. Its solution is 
\begin{equation}
u_i({\bf r}) = -\frac{A}{\mu}\int d^3 r' G_{iz}({\bf r} - {\bf r}')(\partial S_z^2/\partial z), \label{solution-u}
\end{equation}
where
\begin{equation}
G_{ik} = \frac{1}{4\pi}\left[\frac{\delta_{ik}}{r} - \frac{1}{4(1-\sigma)}\frac{\partial^2 r}{\partial r_i \partial r_k}\right] \label{GF-r}
\end{equation}
is the Green function \cite{Elasticity} of Eq.\ (\ref{eq-u-an}). 

If the skyrmion moves along the $x$-axis at a speed $v$ the solution (\ref{solution-u}) must be replaced with  ${\bf u}(x-vt,y,z)$. Its substitution into the first term of Eq.\ (\ref{E-anisotropic}) gives for the kinetic energy, $K.E. = \frac{1}{2 }M_S v^2$, where 
\begin{equation}
M_S = \rho \int d^3 r \left(\frac{\partial {\bf u}}{\partial x}\right)^2
\end{equation}
is the mass of the skyrmion due to spin-phonon coupling. Substitution of Eq.\ (\ref{solution-u}) in the expression for the mass yields
\begin{equation}
M_S  =  \rho \left(\frac{A}{\mu}\right)^2  \int d^3r' \int d^3 r'' F_{zz}({\bf r}' - {\bf r''}) \frac{\partial S_z^2({\bf r}')}{\partial z'}\frac{\partial S_z^2({\bf r}'')}{\partial z''}, 
\label{MassAn}
\end{equation}
where
\begin{equation}
F_{zz}({\bf r}' - {\bf r''})  = \frac{\partial}{\partial x'}\frac{\partial}{\partial x''}\int d^3 r G_{iz}({\bf r}- {\bf r}') G_{iz}({\bf r}- {\bf r}'') . \label{F}
\end{equation}
Using the Fourier transform of the Green function (\ref{GF-r}), 
\begin{equation}
G_{ik}(k) = \frac{1}{k^2}\left[\delta_{ik} - \frac{1}{2(1-\sigma)}\frac{k_ik_k}{k^2}\right],
\end{equation}
$F_{zz}$ can be written as
\begin{equation}
F_{kj}({\bf r}' - {\bf r''})  = \int\frac{d^3k}{(2\pi)^3} e^{-i{\bf k}({\bf r}'-{\bf r}'')} \frac{k_x^2}{k^4}\left[\delta_{kj} - p\frac{k_kk_j}{k^2}\right],
\end{equation}
where 
\begin{equation}
p = \frac{1}{(1-\sigma)}\left[1 - \frac{1}{4(1-\sigma)}\right]
\end{equation}
is the parameter of the elastic theory satisfying $7/16 \leq p \leq 1$.

At this point it suffices to consider a thin-film approximation, $d \ll \lambda$, when one can write 
\begin{equation}
S_z^2({\bf r}) = S_z^2({\bm \rho})d\delta(z)
\end{equation}
with ${\bm \rho} = x{\bf e}_x + y{\bf e}_y$ being the radius-vector in the $xy$ plane of the magnetic layer. Integrating by parts in Eq.\ (\ref{MassAn}) one obtains
\begin{equation}
M_S  =  \rho \left(\frac{A}{\mu}\right)^2 d^2  \int d^2\rho' \int d^2 \rho'' K({\bm \rho}' - {\bm \rho}'') S_z^2({\bm \rho}')S_z^2({\bm \rho}''), 
\label{MassAnThin}
\end{equation}
with
\begin{equation}
K({\bm \rho}) = \int\frac{d^3 k}{(2\pi)^3}  \frac{k_x^2k_z^2}{k^4}\left[1 - p\frac{k_z^2}{k^2}\right]\exp(-ik_x x -ik_y y).
\end{equation}
Changing the order of integration in Eq. (\ref{MassAnThin}) one can write it as
\begin{equation}
M_S  = \rho \left(\frac{A}{\mu}\right)^2 d^2 \int\frac{d^3 k}{(2\pi)^3}  \frac{k_x^2k_z^2}{k^4}\left[1 - p\frac{k_z^2}{k^2}\right]f(k_{\perp}), \label{M-f}
\end{equation}
where $k_{\perp} = \sqrt{k_x^2 + k_y^2}$ and
\begin{equation} 
f(k_{\perp}) = \left|\int d^2\rho \,S_z^2(\rho)\exp(i{\bf k}_{\perp}\cdot{\bm \rho})\right|^2.
\end{equation}
Its independence on $k_z$ allows one to integrate over $k_z$ in Eq. (\ref{M-f}). This leads to 
\begin{equation}
M_S =  \frac{1}{16\pi}\rho \left(\frac{A}{\mu}\right)^2 d^2\left(1-\frac{3}{4}p\right)\int_0^{\infty} d k_{\perp} k_{\perp}^2 f(k_{\perp}).
\end{equation}

We now have to compute $f(k_{\perp})$. Substituting $S_z(\rho) = (\lambda^2 - \rho^2)/(\lambda^2 + \rho^2)$ from Eq.\ (\ref{BP}) into Eq. (\ref{M-f}), we get
\begin{equation}
f(k_{\perp}) =(8\pi)^2{\lambda^4}u^2(k_{\perp}\lambda), \quad u(q) =   -\int_0^{\infty } dr\,r \frac{r^3J_0(qr)}{(1+r^2)^2}, \label{u}
\end{equation}
where $J_0$ is the Bessel function. This gives for the mass
\begin{equation}
M_S =4\pi \rho \left(\frac{A}{\mu}\right)^2 \left(1-\frac{3}{4}p\right)d^2\lambda \int_0^{\infty} d q \, q^2 u^2(q).
\end{equation}
Here $u(q)$ given by Eq.\ (\ref{u}) can be expressed via special functions, which facilitates numerical computation of the integral. The answer yields
\begin{equation}
M_S = 0.787 \left(1-\frac{3}{4}p\right)\left(\frac{A}{\mu}\right)^2 \rho \,d^2\lambda. \label{M-an-final}
\end{equation}
We have double-checked this result by performing a more tedious integration in real space without replacing the layer of thickness $d$  with a $\delta$-function. It produces the same answer with the numeric factor given by
\begin{equation}
c = \frac{1}{16\pi}\int d^2\bar{\rho}' \int d^2 \bar{\rho} \left[S_z^2({\bar{\bm \rho}}' + {\bar{\bm \rho}}) - S_z^2({\bar{\bm \rho}}')\right]^2\frac{2\bar{x}^2 - \bar{y}^2}{\bar{\rho}^5},
\end{equation}
where $\bar{\bm \rho} = {\bm \rho}/\lambda$. This four-dimensional integral reduces to a one-dimensional integral of an awkward elementary function that has a numerical value of $0.785$, very close to the factor in Eq.\ (\ref{M-an-final}). Notice that for the anisotropic magnetoelastic interaction that we have chosen the mass does not depend on the chirality angle $\gamma$. 

To see how general this result is, consider now isotropic magnetoelastic coupling of the form $A u_{ik} S_iS_k$. Repeating the steps of the previous calculation we obtain for the skyrmion mass
\begin{equation}
M_{S}= \rho \left(\frac{A}{\mu}\right)^2 d^2 \int \frac{d^3k}{(2\pi)^3} \frac{k_{x}^2}{k^4} \left[|\mathbf{G}|^2-p \frac{|\mathbf{k}\cdot \mathbf{G}|^2}{k^2}\right],
\end{equation}
where
\begin{equation}
\mathbf{G}= \int dx dy \left(\mathbf{k} \cdot \mathbf{S}\right) \mathbf{S} \exp(-ik_{x}x-ik_{y} y).
\end{equation}
This calculation requires more effort as it involves all three components of the skyrmion spin-field. The final answer reads
\begin{equation}
M_{s}= c(p,\gamma) \rho \left(\frac{A}{\mu}\right)^2 d^2 \lambda,
\end{equation}
with the numerical factor given by,
\begin{eqnarray}
c(p,\gamma) & = & 4.118+0.727 \cos(2\gamma) \\ \nonumber
 & - & p[1.612+0.795\cos(2\gamma)+0.255 \cos(4\gamma)]. \label{c-iso}
\end{eqnarray}
Thus, in general, one should expect the skyrmion mass to depend on both, the elastic properties of the crystal and the chirality of the skyrmion. 

The majority of the materials have $p \sim 1$. At $p = 1$ one obtains from Eq.\ (\ref{c-iso})  $c = 2.186$ for the N\'{e}el skyrmion ($\gamma = 0$) and $c = 2.316$ for the Bloch skyrmion ($\gamma = \pi/2$). Notice that this factor for the anisotropic magnetoelastic coupling at $p = 1$ is $0.197$, which is an order of magnitude smaller. Up to that factor the proportionality of the skyrmion mass to $\rho \left({A}/{\mu}\right)^2 d^2 \lambda$ is robust. The model correctly captures universal scaling of the mass as the square of the strength of the magnetoelastic coupling, the square of the thickness of the ferromagnetic layer, and the first power of the lateral size of the skyrmion. Notice that the proportionality of the skyrmion phonon mass to its size instead of its volume ($V \sim d\lambda^2$) is related to the fact that only spin-field derivatives contribute to the effect. If a thin-wall skyrmion bubble (or a cylindrical domain) of radius $R$ were considered instead, the mass would have been proportional to the area of the wall and would scale linearly with $R$. 

To estimate the magnitude of the effect, notice that the magnetoelastic energy density $A$ is of the relativistic origin (it often comprises a noticeable part of the magnetocrystalline anisotropy), while the shear modulus $\mu$ is of the electrostatic origin arising from the coupling between atoms in a crystal. This allows one to roughly estimate the ratio $A/\mu$ to be in the ballpark of $10^{-4}$. At  $\rho \sim 5\times 10^3$kg/m$^3$ and $d \sim 2$nm it gives $M_S$ of the order of a few electron masses for a skyrmion of size $\lambda \sim 10$nm. However, in materials with high magnetostriction this mass can be significantly greater as it scales as square of the strength of the magnetoelastic coupling. The framework used in this Letter for computing the mass of a skyrmion allows one to develop of a software package for obtaining the skyrmion mass in materials with arbitrary crystal symmetry and arbitrary structure of the magneto-elastic coupling. 

This work has been supported by the grant No. DE-FG02-93ER45487 funded by the U.S. Department of Energy, Office of Science.


\begin{thebibliography}{0}

\bibitem{BelPolJETP75}
A. A. Belavin and A. M. Polyakov, Metastable states of two-dimensional isotropic ferromagnets, Pis'ma Zh. Eksp. Teor. Fiz. \textbf{22}, 503-506 (1975) {[}JETP Lett. \textbf{22}, 245-248 (1975){]}.

\bibitem{WriMerRMR89}
D. C. Wright, and N. D. Mermin, Crystalline liquids: the blue phases, Rev. Mod. Phys. \textbf{61}, 385-433 (1989).

\bibitem{SonKarKivPRB93}
S. L. Sondhi, A. Karlhede, S. A. Kivelson, and E. H. Rezayi, Skyrmions and the crossover from the integer to fractional quantum Hall effect at small Zeeman energies, Phys. Rev. B \textbf{47}, 16419-16426 (1993).

\bibitem{StonePRB93}
M. Stone, Magnus force on skyrmions in ferromagnets and quantum Hall systems, Phys. Rev. B \textbf{53}, 16573-16578 (1996).

\bibitem{Bogdanov94}
A. Bogdanov and A. Hubert, Thermodynamically stable magnetic vortex states in magnetic crystals, Journal of Magnetism and Magnetic Materials {\bf 138}, 255-269 (1994).

\bibitem{YeKimPRL99}
Jinwu Ye, Y. B. Kim, A. J. Millis, B. I. Shraiman, P. Majumdar, and Z. Tesanovic, Berry phase theory of the anomalous Hall effect: Application to colossal magnetoresistance manganites, Phys. Rev. Lett. \textbf{83}, 3737-3740 (1999).

\bibitem{AlkStoNat01}
U. Al'Khawaja, and H. T. C. Stoof, Skyrmions in a ferromagnetic Bose-Einstein condensate, Nature \textbf{411}, 918-920 (2001).

\bibitem{SkyrmePRC58}
T. H. R. Skyrme, A non-linear theory of strong interactions, Proceedings of the Royal Society A \textbf{247}, 260-278 (1958).

\bibitem{Polyakov-book}
A. M. Polyakov, \textit{Gauge Fields and Strings}, Harwood Academic Publishers 1987.

\bibitem{Manton-book}
N. Manton and P. Sutcliffe, \textit{Topological Solitons}, Cambridge University Press 2004.

\bibitem{D1}
E. Braaten and L. Carson, Deuteron as a toroidal skyrmion, Phys. Rev. D \textbf{38}, 3525-3539 (1988).

\bibitem{D2}
W. Y. Crutchfield, N. J. Snyderman, and V. R. Brown, Deuteron in the Skyrme model, Phys. Rev. Lett. \textbf{68}, 1660-1662 (1992).

\bibitem{MS-bubbles} A. P. Malozemoff and J. C. Slonczewski, \textit{Magnetic
Domain Walls in Bubble Materials}, Academic Press 1979.

\bibitem{ODell} T. H. O'Dell, \textit{Ferromagnetodynamics: The Dynamics
of Magnetic Bubbles, Domains, and Domain Walls}, Wiley 1981.

\bibitem{Nagaosa2013}
N. Nagaosa and Y. Tokura, Topological properties and dynamics of magnetic skyrmions, Nature Nanotechnology \textbf{8}, 899-911 (2013).

\bibitem{Zhang2015}
X. Zhang, M. Ezawa, and Y. Zhou, Magnetic skyrmion logic gates: conversion, duplication and merging of skyrmions, Scientific Reports \textbf{5}, 9400-(8) (2015).

\bibitem{Klaui2016}
G. Finocchio, F. B\"{u}ttner, R. Tomasello, M. Carpentieri, and M. Klaui, Magnetic skyrmions: from fundamental to applications, J. Phys. D: Applied Physics \textbf{49}, 423001-(17) (2016).

\bibitem{Leonov-NJP2016}
A. O. Leonov, T. L. Monchesky, N. Romming, A. Kubetzka, A. N. Bogdanov, and R. Wiesendanger, The properties of isolated chiral skyrmions in thin magnetic films, New J. Phys. \textbf{18}, 065003-(16) (2016).

\bibitem{Hoffmann-PhysRep2017}
W. Jiang, G. Chen, K. Liu, J. Zang, S. G. E. te Velthuis, and A. Hoffmann, Skyrmions in magnetic multilayers, Phys. Rep. \textbf{704}, 1-49 (2017).

\bibitem{Fert-Nature2017}
A. Fert, N. Reyren, and V. Cros, Magnetic skyrmions: advances in physics and potential applications, Nature Reviews Materials \textbf{2}, 17031-(15) (2017).

\bibitem{Yu-NanoLet2016}
G. Yu, P. Upadhyaya, Q. Shao, H. Wu, G. Yin, X. Li, C. He, W. Jiang, X. Han, P. K. Amiri, and K. Wang, Room-temperature skyrmion shift device for memory application, Nano Letters \textbf{17}, 261-268 (2016).

\bibitem{Legrand-Nanolet2017}
W. Legrand, D. Maccariello, N. Reyren, K. Garcia, C. Moutafis, C. Moreau-Luchaire, S. Collin, K. Bouzehouane, V. Cros, and A. Fert, Room-temperature current-induced generation and motion of sub-100 nm skyrmions, Nano Letters \textbf{17}, 2703-2712 (2017).

\bibitem{review2020}
C. Back, V. Cros, H. Ebert, K. Everschor-Sitte, A. Fert, M. Garst, Tianping Ma, S. Mankovsky, T. L. Monchesky, M. Mostovoy, N. Nagaosa, S.S.P. Parkin,
C. Peiderer, N. Reyren, A. Rosch, Y. Taguchi, Y. Tokura, K. von Bergmann1, and Jiadong Zang, The 2020 skyrmionics roadmap, arXiv:2021.00026v3, to appear in Journal of Physics D: Applied Physics.

\bibitem{Doring}
W. D$\ddot{\text{o}}$ring, Mikromagnetismus, in Handbuch der Physik, edited by S. Fl$\ddot{\text{u}}$gge, Vol. 18/2, pp. 314-437 (Springer, Berlin, Heidelberg, 1966.) 

\bibitem{Buttner-Nature2015}
F. B$\ddot{\text{u}}$ttner, C. Moutafis, M. Schneider, B. Kr$\ddot{\text{u}}$ger, C. M. G$\ddot{\text{u}}$nther, J. Geilhufe, C. v. Korff Schmising, J. Mohanty, B. Pfau, S. Schaffert, A. Bisig, M. Foerster, T. Schulz, C. A. F. Vaz, J. H. Franken, H. J. M. Swagten, M. Kl$\ddot{\text{u}}$ui, and S. Eisebitt, Dynamics and inertia of skyrmionic spin structures, Nature Physics {\bf 11}, 225 (2015). 

\bibitem{Shiino2017}
T. Shiino, K.-J. Kim, K.-S. Lee, and B.-G. Park, Inertia-driven resonant excitation of a magnetic skyrmion, Nature Scientific Reports {\bf 7}, 13993 (2017). 

\bibitem{Makhfudz-PRL2012}
I. Makhfudz, B. Kruger, and O. Tchernyshyov, Inertia and chiral edge modes of a skyrmion magnetic bubble, Physical Review Letters {\bf 109}, 217201-(4) (2012).

\bibitem{Lin-PRB2017}
S.-Z. Lin, Dynamics and inertia of a skyrmion in chiral magnets and interfaces: A linear response approach based on magnon excitations, Physical Review B {\bf 96}, 014407 (2017).

\bibitem{Psaroudaki-PRX2017}
C. Psaroudaki, S. Hoffman, J. Klinovaja, and D. Loss, Quantum dynamics of skyrmions in chiral magnets, Physical Review X {\bf 7}, 041045-(18) (2017). 

\bibitem{Kravchuk-PRB2018}
V. P. Kravchuk, D. D. Sheka, U. K. R$\ddot{\text{o}}${\ss}ler, J. van den Brink, and Y. Gaididei, Spin eigenmodes of magnetic skyrmions and the problem of the effective skyrmion mass, Physical Review B {\bf 97}, 064403-(10) (2018). 

\bibitem{Li-PRB2018}
Z.-X. Li, C. Wang, Y. Cao, and P. Yan, Edge states in a two-dimensional honeycomb lattice of massive magnetic skyrmions, Physical Review B {\bf 98}, 180407(R)-(6) (2018). 

\bibitem{Liu-MMM2020}
Y. Liu and Z. Liang, Measurement of skyrmion mass by using simple harmonic oscillation, Journal of Magnetism and Magnetic Materials {\bf 500}, 166382 (2020). 

\bibitem{Bogdanov-Nature2006}
U. K. R$\ddot{\text{o}}${\ss}ler, N. Bogdanov, and C. Pfleiderer, Spontaneous skyrmion ground states in magnetic metals, Nature \textbf{442}, 797-801 (2006).

\bibitem{Heinze-Nature2011}
S. Heinze, K. von Bergmann, M. Menzel, J. Brede, A. Kubetzka, R. Wiesendanger, G. Bihlmayer, and S. Blugel, Spontaneous atomic-scale magnetic skyrmion lattice in two dimensions, Nature Physics \textbf{7}, 713-718 (2011).

\bibitem{Boulle-NatNano2016}
O. Boulle, J. Vogel, H. Yang, S. Pizzini, D. de Souza Chaves, A. Locatelli, T. O. Mentes, A. Sala, L. D. Buda-Prejbeanu, O. Klein, M. Belmeguenai, Y. Roussign\'e, A. Stahkevich, S. M. Ch\'erif, L. Aballe, M. Foerster, M. Chshiev, S. Auffret, I. M. Miron, and G. Gaudin, Room-temperature chiral magnetic skyrmions in ultrathin magnetic nanostructures, Nature Nanotechnology \textbf{11}, 449-454 (2016).

\bibitem{DCG-PRB2018}
A. Derras-Chouk, E.M. Chudnovsky, and D.A. Garanin, Quantum collapse of a magnetic skyrmion, Physical Review B {\bf 98}, 024423-(9) (2018). 

\bibitem{Elasticity}
{L. D. Landau and E. M. Lifshitz}, {\it Theory of Elasticity} (Pergamon Press, Oxford, New York, 1986).

\end{thebibliography}
\end{document}